\pdfoutput=1
\documentclass[preprint,authoryear,3p,11pt]{elsarticle}
\usepackage{amssymb}
\usepackage{amsmath}
\usepackage{bbm}
\usepackage{graphicx}
\usepackage{exscale}
\usepackage{hhline}
\usepackage{float}

\biboptions{sort&compress}

\newcommand*\oline[1]{%
  \vbox{%
    \hrule height 0.5pt
    \kern0.25ex
    \hbox{%
      \kern-0.2em
      \ifmmode#1\else\ensuremath{#1}\fi
      \kern+0.0em
    }
  }
}
\newcommand*\ovline[1]{%
  \vbox{%
    \hrule height 0.5pt
    \kern0.25ex
    \hbox{%
      \kern-0.1em
      \ifmmode#1\else\ensuremath{#1}\fi
      \kern+0.0em
    }
  }
}
\newcommand{\bm}[1]{\mbox{\boldmath $#1$}}

\def\beq{\begin{equation}}
\def\best{\begin{equation*}}
\def\eeq{\end{equation}}
\def\eest{\end{equation*}}
\def\ni{\noindent}
\def\tin{t\in[0,\infty)}

\def\tintau{t\in[0,\tau]}
\def\mf{\mathcal{F}}
\def\ws{\widehat{S}}

\begin{document}

\title{Loading Pricing of Catastrophe Bonds and Other Long-Dated, Insurance-Type Contracts}
\author[UTS,UCT]{Eckhard Platen}
\author[UCT]{David Taylor}
\address[UTS]{Finance Discipline Group and School of Mathematical and Physical Sciences, University of Technology, Sydney, PO Box 123, Broadway, NSW, 2007, Australia}
\address[UCT]{Dept. of Actuarial Science and the African Collaboration for Quantitative Finance \& Risk Research, University of Cape Town, Private Bag X3, Rondebosch 7701, South Africa}

\begin{abstract}
Catastrophe risk is a major threat faced by individuals, companies, and entire economies. Catastrophe (CAT) bonds have emerged as a method to offset this risk and a corresponding literature has developed that attempts to provide a market-consistent pricing methodology for these and other long-dated, insurance-type contracts. This paper aims to unify and generalize several of the widely-used pricing approaches for long-dated contracts with a focus on stylized CAT bonds and market-consistent valuation. It proposes a loading pricing concept that combines the theoretically possible minimal price of a contract with its formally obtained risk neutral price, without creating economically meaningful arbitrage. A loading degree controls how much influence the formally obtained risk neutral price has on the market price. A key finding is that this loading degree has to be constant for a minimally fluctuating contract, and is an important, measurable characteristic for prices of long-dated contracts. Loading pricing allows long-dated, insurance-type contracts to be priced less expensively and with higher return on investment than under classical pricing approaches. Loading pricing enables insurance companies to accumulate systematically reserves needed to manage its risk of ruin in a market consistent manner.
\end{abstract}

\maketitle
\vfill


\pagebreak
\sloppypar
\setcounter{footnote}{1}

\section{Introduction}
\ni Long-dated contracts are essential for risk management and investment in insurance, pension funds and banking. Long-dated CAT bonds are a typical example of such contracts. This paper takes the view that classical no-arbitrage (risk neutral) pricing theory yields prices for long-dated contracts that may appear to many investors as being too expensive and yielding returns on investment that are too low. This could potentially explain why CAT bonds have only gradually been gaining popularity. A more attractive pricing approach than is currently used may supply the stimulus required for CAT bonds to gain the market share needed to cover most economic losses arising from catastrophes.\\

\ni To give another indication of why a new level of pricing theory may be needed, note that under the classical pricing theory for generating the payoff of one unit of the savings account at maturity, one is advised by the Law of One Price to invest simply in one unit of the savings account. However, this is not consistent with commonly given financial planning advice: Those who are young should invest in risky securities and those who are closer to retirement should invest more in riskless securities such as the savings account. Despite the prevalence of this advice, it is unsupported by classical asset pricing theory. Nevertheless, this financial planning advice is common because it has been successful for many generations. Due to risky securities being held at the beginning of the life-cycle, the strategy provides a higher average return on investment than investing purely in riskless securities throughout a person's life.\\

\ni The fact that the advice is unsupported by classical asset pricing theory matters. It matters because classical asset pricing theory makes the current price of a long-dated security with a riskless payout higher than it needs to be. It does so because the low return on investment of the riskless security determines the high price for the targeted riskless payout. Thus, classical asset pricing theory is incompatible with typical practice in financial planning, but impacts practice by being used to set prices that are higher than necessary.\\

\ni There is an urgent need to consolidate this practice under a generalized pricing concept. In particular, for developed economies with low or even negative interest rates, the demand to achieve higher returns on investment for pension funds, life insurance contracts, green infrastructure funding, and long-dated insurance contracts, such as CAT bonds, has become extreme.\\

\ni By avoiding the restrictive assumptions of the classical asset pricing approach, the current paper proposes the new concept of loading pricing, where the theoretically possible minimal price is combined with a formally obtained risk neutral price. A loading degree determines how much the formally obtained risk neutral price influences the loading price. We will show that there is no economically meaningful arbitrage if the loading degree is a local martingale and orthogonal to already-traded uncertainty. The loading degree represents a new and important, observable quantity for long-dated contracts. It enables the market-consistent design of contracts such as CAT bonds with higher returns on investment than previously thought. This makes these products less expensive and more attractive to long-term investors. Our new concept of loading pricing is widely applicable. We motivate it throughout the paper by the challenge of market-consistent valuation of long-dated CAT bonds.\\

\ni The outline of the paper is as follows: In Section 2 we give a brief summary of the pricing of CAT bonds and other long-dated, insurance-type products. Section 3 provides an introduction to the benchmark approach that underpins our general modeling framework. The new concept of loading pricing is defined in Section 4. Formally obtained risk neutral prices are derived in Section 5. Examples for stylized CAT bonds illustrate the new pricing concept in Section 6. Section 7 discusses hedging, and Sections 8 and 9 risk minimization.

\section{Brief Review of Pricing CAT Bonds and Long-Dated, Insurance-Type Contracts}
\ni Since the early 1990s, insurance and reinsurance companies, in conjunction with banks, have been devising innovative ways to transfer catastrophe risk to the capital markets; motivated in part by the size of the daily fluctuations of international capital markets when compared with natural disaster (re)insurance claims. The 10-year, inflation-adjusted average of total economic losses from natural catastrophes and man-made disasters stood at US \$192 billion in 2015 (\cite{re2016natural}), a value easily swamped by the daily volatility of global capital. Interconnected transfer mechanisms emerged to facilitate this transfer, all of which were originally based on a constructed loss index. These mechanisms included exchange-traded catastrophe futures and options, CAT bonds, and, most recently, over-the-counter catastrophe swaps. Initially, pricing and hedging efforts concentrated on catastrophe futures and options since their values were referenced to an exchange-quoted index. Three attempts were made to quote and trade these futures and options: firstly, between 1992 and 1995 on the Chicago Board of Trade (CBOT), which referenced its own loss index; secondly, between 1995 and 2000, which referenced an index constructed by the Property Claims Service (PCS); and finally, between 2007 and 2010 on the New York Mercantile Exchange (NYMEX), which settled against the Re-Ex loss index (in turn constructed from data supplied by PCS). Insurers and investors showed little interest in all three attempts, which led to a thinly-traded market, blamed, in part, on a lack of relevant expertise by the exchanges. According to~\cite{cummins2008cat}, counterparty credit risk, moral hazard, disruption of long-term relationships with reinsurers, and the possibility of excessive basis risk further impaired liquidity. As suggested in the introduction, one may also have to blame the low return on investment that results when pricing these securities according to classical asset pricing theory. Finally, the non-traded nature of the underlying index made hedging all but impossible for market makers.\\

\noindent Gradually, CAT futures and options gave way in popularity to CAT bonds, which have emerged as the leading method to offset catastrophe risk. The earliest CAT bond issuance by Svensk Exportkredit was in 1984. Recent data show outstanding issuance growing from US \$3.7 billion to US \$23.2 billion between 2004 and 2016, which covers over 10\% of the average economic losses due to catastrophes; see~\cite{willis2016}. CAT bonds are simple in design, and the stylized CAT bonds we will use later on for illustration are chosen to be extremely simple. Their price is contingent on insurance claims from natural catastrophes, and an enhanced yield is offered to compensate for the partial or complete loss of principal and/or coupons. The contingency value may be either the insurer's own losses (thus eliminating basis risk) or industry-wide losses, as reflected by an agreed index. A description of the design, features, former and current pricing for CAT futures, options and bonds, insurance products, and reinsurance can be found in e.g.~\cite{litzenberger1996assessing},~\cite{canabarro2000analyzing},~\cite{cox2000economic},~\cite{froot2001market},~\cite{cummins2008cat},~\cite{cummins2009convergence},~\cite{cummins2012geneva},~\cite{huynh2013review},
 ~\cite{willis2016},~\cite{gibson2014reinsurance}, and~\cite{gurtler2014impact}.\\

\noindent Four distinct approaches to CAT bond pricing appear to have been proposed to date.~\cite{embrechts1997pricing} produced a general pricing framework based on utility maximization, which included a discussion of market completeness and a caution about using standard no-arbitrage pricing methods.~\cite{aase1999} selected a partial equilibrium framework with constant absolute risk aversion in which CAT risk was regarded as systemic.~\cite{cox2000} also employed equilibrium pricing theory and included a time separable utility function, while~\cite{christensen2000} introduced an exponential utility model  within a similar framework.~\cite{aase2001} extended his earlier work to a competitive equilibrium approach with constant relative risk aversion. Following on from the foundational work of~\cite{buhlmann1984, buhlmann1985}, actuarial pricing methods were employed in~\cite{lane2000} and~\cite{lane2008}, where the link between securitization of CAT risk and reinsurance is clearly noted.\\

\noindent Quite rapidly the strands of theory coalesced around some preference-free, no-arbitrage framework. The motivation for such an approach stems from the incomplete market framework considered in~\cite{merton1976}. In these models, natural catastrophes have been treated as idiosyncratic risks that can be (almost) fully diversified. Subsequent pricing methods have assumed that the risk neutral probability measure has been predetermined, usually by assuming that CAT risk is orthogonal to market risk, thus avoiding complicated changes of measure. This assumption is supported by the empirical studies of~\cite{hoyt1999} and~\cite{cummins2009convergence}, and disputed in~\cite{carayannopoulos2015diversification} and~\cite{gurtler2014impact}. Under an assumed risk neutral probability measure, stochastic processes and distributions used to price CAT bonds retain the same characteristics as under the real world probability measure. Unfortunately, calibration of these models seems all but impossible due to a lack of data. Many of the models simply propose a plausible index process, and then employ simulation in order to price and hedge CAT derivatives.\\

\noindent The no-arbitrage approach to reinsurance contracts was introduced in~\cite{sondermann1991}. The first to apply risk neutral methods to CAT futures were~\cite{cummins1994, cummins1995}.~\cite{chang1996} developed a CAT futures options model based on a stochastic time change.~\cite{Baryshnikov1998} developed a classical no-arbitrage model for CAT bonds using a compound Poisson process and a change-of-measure argument, without considering market completeness. This model was extended in~\cite{burnecki2003}, and used by~\cite{dassios2003} to model stop-loss reinsurance contracts and CAT derivatives. Thereafter, various extensions of these initial works were applied to CAT futures, options, bonds, and swaps; see e.g.~\cite{geman1997},~\cite{louberge1999},~\cite{lee2002},~\cite{bakshi2002},~\cite{muermann2002, muermann2003, muermann2008},~\cite{schmidli2003},~\cite{vaugirard2003a, vaugirard2003b, vaugirard2004},~\cite{albrecher2004qmc},~\cite{jaimungal2006},~\cite{lee2007},~\cite{zimbidis2007},~\cite{biagini2008},~\cite{egami2008},~\cite{lin2009},~\cite{linetal2009},~\cite{jarrow2010},
~\cite{braun2011},~\cite{chang2011},~\cite{ma2013pricing},~\cite{nowak2013pricing}, and ~\cite{jaimungal2014}.\\

\noindent A literature has been growing that introduces market models that cannot be accommodated by classical no-arbitrage pricing theory. An equivalent risk neutral probability measure may not exist for many realistic models, described e.g. in~\cite{DelbaenSchachermayer95b},~\cite{LoewensteinWillard00},~\cite{fernholz2002stochastic},~\cite{Platen02, Platen06},~\cite{fernholz2005diversity},~\cite{PlatenHeath06},~\cite{JarrowProtterShimbo10},~\cite{KaratzasKardaras07},~\cite{HestonLoewensteinWillard07},~\cite{ChristensenLarsen07},~\cite{FernholzKaratzas09},~\cite{biagini2009local},
~\cite{ GalessoRunggaldier10},~\cite{FernholzKaratzas10},~\cite{DavisLleo11},~\cite{biagini2013evaluating},~\cite{Hulley10},~\cite{Ruf11},~\cite{biagini2014local}, and~\cite{baldeaux2015pricing}.~\cite{heath2002consistent} demonstrated that pricing and hedging is still possible in models that lead beyond the classical asset pricing framework. The work in~\cite{Platen02} and~\cite{PlatenHeath06} suggests that, for realistic long-term modeling, one has to go beyond classical risk neutral pricing. The benchmark approach, established in~\cite{Platen02, Platen06} and presented in~\cite{PlatenHeath06}, provides a general modeling framework for pricing and hedging in incomplete markets. It can accommodate a wide range of models and pricing approaches beyond those covered by the classical risk neutral pricing approach.\\

\noindent Under the benchmark approach, asset prices are modeled under the real world probability measure and are typically denominated in units of the num\'eraire portfolio (NP); see ~\cite{Long90}. The NP is called the benchmark and any price denominated in units of the NP is the respective benchmarked price. This portfolio can be shown to yield, almost surely, the highest portfolio path in the long run and to maximize expected log-utility. This means that it is equivalent to the growth optimal portfolio, discovered in~\cite{kelly1956new}. In~\cite{Platen02} and~\cite{buhlmann2003discrete} it was pointed out that the restrictive assumption on the existence of an equivalent risk neutral probability measure can be avoided under the benchmark approach. For instance, one can perform, so called, real world pricing instead of risk neutral pricing, taking expectations using the real world probability measure and using the NP as num\'eraire. If an equivalent risk neutral probability measure exists for a given market model, then risk neutral pricing turns out to be equivalent to real world pricing. In a complete market, the benchmarked savings account, i.e. the savings account denominated in units of the NP, is the Radon-Nikodym derivative of the putative risk neutral probability measure. When this process is not a true martingale, the assumptions of the classical risk neutral pricing approach break down. The benchmark approach still operates in this case. In particular, real world pricing can be performed. A range of other price processes can then be formed without generating any economically meaningful arbitrage. As we will show later on, the real world price characterizes the theoretically possible minimal price, and there is significant freedom in pricing under the benchmark approach.\\

\noindent The hedging of long-dated, insurance-type contracts, which invariably involve not-fully-replicable contingent claims, has been a challenging task. Several approaches that deal with this problem also appear in the CAT bond literature, to which we referred earlier. Strategies that aim to deliver a not-fully-replicable contingent claim usually generate a fluctuating profit-and-loss process. The classical risk minimization approach of~\cite{FollmerSondermann86} (further developed in~\cite{FollmerSchweizer89},~\cite{Schweizer91, Schweizer99}, and~\cite{moller2001risk}), minimizes fluctuations of discounted profit-and-loss processes by using a quadratic criterion under an assumed minimal equivalent martingale measure. Under this assumed risk neutral probability measure the resulting discounted profit-and-loss process forms a local martingale, orthogonal to discounted traded wealth. This pricing and hedging approach requires the respective Radon-Nikodym derivative, i.e. the benchmarked savings account, to be a true martingale. This is a strong assumption that has been challenged as being unrealistic; see e.g.~\cite{baldeaux2015pricing},~\cite{PlatenHeath06} and~\cite{fergusson2014}. Furthermore, it requires second moment properties, and is not easy to implement; see~\cite{heath2001comparison}. To overcome these shortcomings,~\cite{du2016benchmarked} proposed the concept of benchmarked risk minimization, which yields the theoretically possible minimal price for a not-fully-replicable claim, where, under the real world probability measure, the benchmarked profit-and-loss is a local martingale and orthogonal to benchmarked traded wealth. When classical risk neutral assumptions are satisfied it recovers the risk neutral price obtained under the minimal equivalent martingale measure; see~\cite{Schweizer99}. However, unlike classical risk minimization; see~\cite{moller2001risk}, the resulting hedging strategy takes evolving information about the non-hedgeable part of a benchmarked contingent claim into account. Real world pricing can also be obtained via utility indifference pricing, as shown in Chapter 12 of~\cite{PlatenHeath06}. As outlined in~\cite{du2016benchmarked}, the real world price provides a lower bound for possible prices. It can be evaluated once the NP and the benchmarked contingent claim have been identified and modeled. \\

\ni Insurance companies need to control their probability of ruin and have to satisfy regulatory capital requirements. The risk premium encapsulated in market prices is variously referred to in the literature as safety, security, or contingency loading; see e.g. ~\cite{buhlmann1970mathematical},~\cite{ buhlmann1984, buhlmann1985},~\cite{christensen2000},~\cite{lane2000},~\cite{buhlmann2005course}, and~\cite{buchwalder2007valuation}. These are necessary loadings to manage the risk of ruin for the insurer. It has been difficult to propose a theoretically consistent pricing concept which incorporates a loading that facilitates the accumulation of reserves in a systematic manner without significantly overpricing particular products. Furthermore, there already exists a wide range of prices that can be applied to instruments that could hedge part of an insurance-type contract, or price a similar type of contract. Finally, there is a challenge to propose a ``market-consistent" valuation concept, which needs to include a loading degree from the perspective of the theoretically possible minimal price to avoid bankruptcy of the issuer.\\

\noindent The current paper makes the following contribution to the literature: It points out that a wide range of pricing rules is theoretically possible for not-fully-replicable contingent claims without creating any economically meaningful arbitrage. Using the benchmark approach, it identifies the theoretically possible minimal price (hereafter referred to as just the ``minimal price"), as well as a formally obtained risk neutral price (hereafter referred to as just the ``risk neutral price"). Using the new concept of a loading degree it combines these two prices in a manner that is consistent with the NP and other market prices. The market then has the freedom to form a loading price by choosing a respective loading degree process. We will show that the loading degree process needs to have certain properties, but can reflect behavioral effects in response to recent catastrophes or market downturns. When pricing and hedging are performed under benchmarked risk minimization, the loading degree has to be constant over time for a given contingent claim. This is an important insight from a modeling perspective hinting at ideally only slightly changing loading degrees, whereas random loading degrees introduce extra uncertainty in the market. It also provides an intuitive understanding of the ramifications of the proposed loading pricing concept. Although we motivate and illustrate the proposed loading pricing in the context of pricing and hedging long-dated CAT bonds, we emphasize that this market-consistent valuation methodology applies to many other contingencies in insurance, pension funds, financial planning, and finance.

\section{Introduction to the Benchmark Approach}
\noindent We model a semimartingale market on a filtered probability space $(\Omega,\mathcal{F},\underline{\mathcal{F}},P)$ with filtration $\underline{\mathcal{F}} = (\mathcal{F}_t)_{t\geq 0}$, satisfying the usual conditions. A central notion we will frequently use is that of a local martingale; see~\cite{Protter05}. Intuitively, one can say that over an infinitesimal time period the current value of a local martingale is the best forecast of its instantaneous future value. A special case of a local martingale is that of a martingale, where its current value is the best forecast of all its future values. Obviously, local martingale and martingale processes depend on the underlying probability measure $P$ and the available information at time $t$ modeled via $\mathcal{F}_t$, $t\in[0,\infty)$. We make the following key assumption of the benchmark approach; see~\cite{PlatenHeath06} and~\cite{du2016benchmarked}:\\

\noindent {\bf Assumption 1:} {\it There exists a self-financing, strictly positive portfolio, called the num\'eraire portfolio (NP), which is characterized by the property that it makes all self-financing portfolios of traded securities, including cum-dividend stocks, savings accounts and derivatives, local martingales when these are benchmarked, i.e., denominated in units of the NP.}\\

\noindent This leads us into the general modeling world of the benchmark approach, where we refer to the NP as the benchmark, and to prices denominated in units of the benchmark as benchmarked prices. Assumption 1 is an extremely weak condition. In a general semimartingale market setting when assuming the existence of the NP, it has been shown in ~\cite{ChristensenLarsen07} that all benchmarked, self-financing portfolios are local martingales. With Assumption 1, we also assume the existence of the NP, which means that at any finite time it has some finite value, and does not explode. Such explosion would have to be interpreted as the presence of some economically meaningful arbitrage. As shown in~\cite{PlatenHeath06}, the NP is the strictly positive portfolio that almost surely outperforms with its path all other strictly positive portfolios in the long run.\\

\noindent Under classical risk neutral, no-arbitrage pricing assumptions and in a complete market setting, it was first observed in~\cite{Long90} that, by using the NP as num\'eraire, benchmarked securities and benchmarked risk neutral derivative prices become martingales under the real world probability measure $P$. In our much wider modeling world, we allow benchmarked prices to form local martingales. Thus, we can model benchmarked prices and securities that may not be true martingales. This opens the possibility of new phenomena to be exploited by risk management strategies. As we will explain later, these could potentially make long-dated financial contracts, such as CAT bonds, less expensive than is currently possible under classical assumptions.\\

\noindent Note that market participants can hold securities long or short in their portfolios. Since the benchmarked values of traded securities are local martingales, so are the benchmarked total portfolio values of market participants. This means that all these value processes are consistent with the  NP through the local martingale property of their benchmarked values. This also means that there does not exist any strategy that allows for the construction of a strictly positive portfolio that outperforms the NP in the long run. In particular, it automatically excludes economically meaningful arbitrage in the sense that there might exist a strictly positive portfolio that explodes in finite time. This guides us to our second assumption:\\

\noindent {\bf Assumption 2:} {\it If one adds a new price process to the market, then this has to be consistent with the existing NP such that its benchmarked value is a local martingale.}\\

\ni Consequently, the extended market then has the same NP as before, and the new price process remains consistent with the existing NP. We will see that the above local martingale property, required in Assumption 2 for new benchmarked prices, is a natural condition. It also allows us to establish properties of potential new pricing rules. \\

\ni By Fatou's lemma; see e.g.~\cite{Protter05}, nonnegative local martingales are supermartingales. The current value of a nonnegative supermartigale is greater than or equal to its expected future values. Since this is a fundamental property of benchmarked, nonnegative securities, we summarize it as follows:\\

\noindent {\bf Corollary 1:} {\it Benchmarked, nonnegative price processes are supermartingales.}\\

\ni For a bounded stopping time $\tau\in[0,\infty)$, we call an $\mathcal{F}_{\tau}-$measureable, nonnegative payoff $\widehat{H}_{\tau}$ at time $\tau$ a benchmarked contingent claim if
\begin{equation*}\label{eq:1}
E(\widehat{H}_{\tau}) < \infty.
\end{equation*}
By Corollary 1, we know that a benchmarked, nonnegative price process $\widehat{R}^{\widehat{H}_\tau} = \{\widehat{R}^{\widehat{H}_\tau}(t),\tintau\}$, valuing the benchmarked contingent claim $\widehat{H}_{\tau}$, has to be a supermartingale. This means that
\beq\label{eq:1'}
\widehat{R}^{\widehat{H}_\tau}(t)\geq E(\widehat{R}^{\widehat{H}_\tau}(s)|\mf_t)
\eeq
for $0\leq t\leq s\leq\tau<\infty$. Under any reasonable pricing rule at maturity $\tau$, the benchmarked contingent claim is observable and equal to its benchmarked price. According to Lemma 11.1 in~\cite{du2016benchmarked}, the minimal, nonnegative supermartingale that coincides at a given time $\tau$ with a given nonnegative, $\mathcal{F}_{\tau}-$measureable random variable $\widehat{H}_{\tau}$ is the martingale $\widehat{V}^{\widehat{H}_\tau} = \{\widehat{V}^{\widehat{H}_\tau}(t),t\in[0,\tau]\}$ with $\widehat{V}^{\widehat{H}_\tau}(\tau) = \widehat{H}_{\tau}$ a.s. This leads us to the following result:\\

\noindent {\bf Corollary 2:} {\it For a benchmarked contingent claim $\widehat{H}_{\tau}$, delivered at time $\tau\in[0,\infty)$, the benchmarked, minimal price process $\widehat{V}^{\widehat{H}_\tau}$ is determined by the real world pricing formula
\begin{equation}\label{eq:2}
\widehat{V}^{\widehat{H}_\tau}(t) = E\left(\widehat{H}_\tau|\mathcal{F}_{t}\right)
\end{equation}
for $0\leq t\leq \tau<\infty.$}\\

\noindent The minimal price provides us with a lower bound for price processes consistent with the NP. In~\cite{du2016benchmarked} it has been shown that this is the price that emerges in a competitive market when financial institutions eliminate all hedgeable uncertainty and fully diversify all nonhedgeable risks in their trading book without any fees being charged. The existence of the above lower bound does not force the market to agree on this minimal price. To avoid bankruptcy of issuing companies, the market needs to agree on some higher price. We will propose later on the new concept of loading pricing, where this can be done consistently.\\

\ni To allow us to consider pricing and hedging systematically, we introduce in our market $d\in \{2,3,\ldots\}$ primary security accounts with nonnegative values $S_t^1,S_t^2,\ldots,S_t^d$ at time $t\in[0,\infty)$, denominated in units of the domestic currency. These may include liquid cum-dividend stocks, savings accounts in given currencies, derivative price processes, and traded price processes for various contracts. For convenience, we denote by $S_t^1$ the time $t$ value of the savings account in the domestic currency, which, in practice, is approximated by a rolled-over, short-term bond account. The central building block of our market is the NP, which we denote by $N_t$ at time $t\in[0,\infty)$ and use as num\'eraire or benchmark. The benchmarked value of the $j$th primary security account $S_t^j$, which means its value denominated in units of $N_t$, is then denoted by
\begin{equation*}\label{eq:3}
\widehat{S}^j_t = \frac{S^j_t}{N_t}
\end{equation*}
for $t\in[0,\infty)$ and $j\in\{1,2,\ldots,d\}$. Consequently, the process $\mathbf{\widehat{S}} = \{\mathbf{\widehat{S}}_t = (\widehat{S}^1_t,\ldots,\widehat{S}^d_t)^\top,t\in[0,\infty)\}$ forms, according to Assumption 1, a vector local martingale. The benchmarked value $\widehat{S}^\delta_t$ at time $t\in[0,\infty)$ of a portfolio that holds $\delta^j_t$ units of the $j$th primary security account, $j\in\{1,2,\ldots,d\}$, at time $t$ is given by
\begin{equation*}\label{eq:4}
\widehat{S}^\delta_t = \sum^d_{j=1}\delta^j_t\widehat{S}^j_t = \bm\delta^\top_t\widehat{\mathbf{S}}_t.
\end{equation*}
A strategy $\bm\delta = \{\delta_t = (\delta^1_t,\delta^2_t,\ldots,\delta^d_t)^\top,t\in[0,\infty)\}$ is called self-financing if the process $\bm\delta$ is predictable and one has
\begin{equation}\label{eq:5}
\widehat{S}^\delta_t = \widehat{S}^\delta_0 + \int^t_0\bm\delta^\top_sd\widehat{\bm{S}}_s
\end{equation}
for $t\in[0,\infty)$. The stochastic integral in (\ref{eq:5}) is assumed to be a vector It\^o integral; see e.g.~\cite{du2016benchmarked}. By Assumptions 1 and 2, all benchmarked self-financing portfolios are local martingales. The changes in their value are only due to changes in value of primary security accounts and not to changes in the strategy.\\

\noindent Let us emphasize that in our benchmark framework we have less restrictive assumptions than those imposed in classical finance. The latter have been  generally characterized in~\cite{DelbaenSchachermayer98} to allow the application of the risk neutral pricing rule via measure transformation. In Assumptions 1 and 2, we are demanding only the existence of the NP for all traded and potential new securities. In~\cite{KaratzasKardaras07}, necessary and sufficient conditions have been derived for semimartingale markets that ensure the existence of the NP. The required finiteness of the NP at finite times is a very clear and economically meaningful no-arbitrage condition. Since we are working beyond classical assumptions, this also means that we can model various phenomena and forms of classical arbitrage in our general setting that have been excluded so far under the classical approach.

\section{Loading Pricing}
\ni As shown in~\cite{du2016benchmarked}, the real world pricing formula (\ref{eq:2}) introduces the minimal price for a given, potentially not-fully-replicable, benchmarked contingent claim $\widehat{H}_\tau$. We have already emphasized that a wide range of price processes can be theoretically chosen to price the claim $\widehat{H}_\tau$ consistently with the NP. Furthermore, we need to expect that some loading will be required by the issuer of a not-fully-replicable claim in order to build up and maintain capital reserves. These reserves are needed to reduce and manage the probability of ruin of the issuer; see e.g. ~\cite{buhlmann1970mathematical},~\cite{ buhlmann1984, buhlmann1985},~\cite{christensen2000},~\cite{lane2000}, and~\cite{buchwalder2007valuation}. The methods of pricing insurance contracts with some incorporated loading vary significantly and are often opaque. We aim to generalize at least the formally applied risk neutral pricing rule and the aforementioned real world pricing rule. Risk neutral pricing is widely practiced in the market and real world pricing provides an objective minimal price.\\

\ni According to Assumption 2, we assume that we have a pricing rule for a nonnegative, benchmarked contingent claim $\widehat{H}_\tau$ that yields a price $R^{\widehat{H}_\tau}(t)$ at time $t$, which, when benchmarked and denoted by $\widehat{R}^{\widehat{H}_\tau}(t) = R^{\widehat{H}_\tau}(t)/N_t$, forms a local martingale. \\

\ni A range of prices could be theoretically possible for a given contingent claim above the minimal price $V^{\widehat{H}_\tau}(t)$ at time $t$. By Corollary 2, the fact that $\widehat{V}^{\widehat{H}_\tau}(t) = V^{\widehat{H}_\tau}(t)/N_t$ forms the minimal possible supermartingale that delivers the benchmarked contingent claim $\widehat{H}_\tau$ leads to the following result:\\

\ni  {\bf Corollary 3:} {\it For a nonnegative, benchmarked contingent claim $\widehat{H}_\tau$ its minimal price $V^{\widehat{H}_\tau}(t)$ and a price $R^{\widehat{H}_\tau}(t)$ satisfying Assumption 2 are related by the inequality
\begin{equation*}\label{eq:6}
V^{\widehat{H}_\tau}(t)\leq R^{\widehat{H}_\tau}(t)
\end{equation*}
for all $t\in[0,\tau]$.}\\

\ni This inequality shows that, if the two prices are not equal, then there is scope to form a family of loading prices that range from the minimal price $V^{\widehat{H}_\tau}(t)$ to a more expensive price $R^{\widehat{H}_\tau}(t)$. Currently, many prices in the market for short-dated, hedgeable claims are most likely set by using classical risk neutral pricing. To acknowledge this fact, we will set $R^{\widehat{H}_\tau}(t)$ equal to the risk neutral price, which will be specified in the next section. Now, for a nonnegative, benchmarked contingent claim $\widehat{H}_\tau$, we introduce a nonnegative, predictable loading degree process $L^{\widehat{H}_\tau} = \{L^{\widehat{H}_\tau}_t, t\in[0,\tau]\}$ that is a semimartingale, left continuous with right hand limits. We form the loading price $B^{\widehat{H}_\tau}(t)$ at time $t\in[0,\tau]$ by the equation
\begin{equation}\label{eq:7}
B^{\widehat{H}_\tau}(t) = L^{\widehat{H}_\tau}_t R^{\widehat{H}_\tau}(t) + (1-L^{\widehat{H}_\tau}_t)V^{\widehat{H}_\tau}(t).
\end{equation}
By Corollary 3, we arrive directly at the following conclusion:\\

\ni  {\bf Corollary 4:} {\it For a nonnegative, benchmarked contingent claim $\widehat{H}_\tau$ and loading degree process $L^{\widehat{H}_\tau}$ the loading price $B^{\widehat{H}_\tau}(t)$ satisfies the inequality
\begin{equation*}\label{eq:8}
V^{\widehat{H}_\tau}(t)\leq B^{\widehat{H}_\tau}(t)
\end{equation*}
for $t\in[0,\tau]$.}\\

\ni Recall that $V^{\widehat{H}_\tau}(t)$ is the minimal price. From (\ref{eq:7}) we obtain the loading degree from the (potentially traded) loading price, the minimal price and the risk neutral price via the formula
\begin{equation}\label{eq:9}
L^{\widehat{H}_\tau}_t = \frac{B^{\widehat{H}_\tau}(t) - V^{\widehat{H}_\tau}(t)}{R^{\widehat{H}_\tau}(t) - V^{\widehat{H}_\tau}(t)}
\end{equation}
for $t\in[0,\tau]$, as long as $R^{\widehat{H}_\tau}(t) > V^{\widehat{H}_\tau}(t)$. This makes the loading degree observable, assuming that we have a model that provides the minimal and risk neutral prices. In the case when $R^{\widehat{H}_\tau}(t) = V^{\widehat{H}_\tau}(t)$, we set $L^{\widehat{H}_\tau}_t = 1$ in (\ref{eq:9}). We can rewrite the benchmarked loading price in (\ref{eq:7}) as
\begin{eqnarray}\label{eq:9'}
\widehat{B}^{\widehat{H}_\tau}(t) &=& L^{\widehat{H}_\tau}_t \widehat{R}^{\widehat{H}_\tau}(t) + (1-L^{\widehat{H}_\tau}_t)\widehat{V}^{\widehat{H}_\tau}(t) \nonumber \\
&=& \widehat{V}^{\widehat{H}_\tau}(t) + L^{\widehat{H}_\tau}_t(\widehat{R}^{\widehat{H}_\tau}(t) - \widehat{V}^{\widehat{H}_\tau}(t)).
\end{eqnarray}

\ni To satisfy Assumption 2, it is not enough to note that $\widehat{R}^{\widehat{H}_\tau}$ and $\widehat{V}^{\widehat{H}_\tau}$ are local martingales. We also need $\widehat{B}^{\widehat{H}_\tau}$ to form a local martingale.\\

\ni To facilitate this, the following notion of orthogonality; see e.g.~\cite{du2016benchmarked}, will be important: Two local martingales are said to be orthogonal if their product forms a local martingale. Since $\widehat{V}^{\widehat{H}_\tau}$ in (\ref{eq:9'}) is a local martingale, and the sum of local martingales is itself a local martingale, we require here that the product $L^{\widehat{H}_\tau}_t(\widehat{R}^{\widehat{H}_\tau}(t) - \widehat{V}^{\widehat{H}_\tau}(t))$ forms a local martingale. By using the above notion of orthogonality, we directly obtain the following result:\\

\ni {\bf Corollary 5:} {\it For a given benchmarked contingent claim $\widehat{H}_\tau$, the benchmarked loading price process $\widehat{B}^{\widehat{H}_\tau}$ is a local martingale if the respective loading degree $L^{\widehat{H}_\tau}_t$ forms a local martingale, orthogonal to $\widehat{R}^{\widehat{H}_\tau}(t) - \widehat{V}^{\widehat{H}_\tau}(t)$.}\\

\ni Note that this result is rather general and makes the concept of loading pricing highly tractable. One has only to check for the local martingale condition, i.e. for zero drifts in the respective stochastic differential equations under a given model. The simplest loading degree process would be a constant one. We will see later on that a constant loading degree coincides with some kind of optimal choice with respect to risk minimization because it avoids the introduction of an additional source of uncertainty through the loading degree in the market. In general, $L^{\widehat{H}_\tau}_t$ need not be a constant, and is, according to Corollary 5, a local martingale that fluctuates (due to orthogonality) very differently to $R^{\widehat{H}_\tau}(t) - V^{\widehat{H}_\tau}(t)$. This means that such a fluctuating loading degree adds a new source of uncertainty to the market, which could be interpreted as the uncertainty generated by discovering the market price, e.g. via supply-and-demand or behavioral features. We note that there is significant freedom in pricing long-dated contingent claims using the concept of loading pricing. Such flexibility is not available under the classical approach, which in effect sets the loading degree to one. Later on we will see that it is possible to set the loading degree to a process with values less than one, which leads to a form of market-consistent valuation that allows less expensive contract prices. Of course, the loading degree process could also be greater than one, but not less than zero.\\

\section{Formally Obtained Risk Neutral Prices}
\noindent In our modeling framework of the benchmark approach we cover a wide range of models that do not have a benchmarked savings account process $\ws^1$ that forms a true martingale. In such a situation, and when we have a complete market, the Radon-Nikodym derivative $\Lambda^Q_t = \frac{\ws^1_t}{\ws^1_0} = \left.\frac{dQ}{dP}\right|_{\mathcal{F}_t}$ of the putative risk neutral probability measure $Q$ forms a strict local martingale and there does not exist an equivalent risk neutral probability measure. In this case one cannot apply the classical theory to form a theoretical risk neutral price for a given contingent claim because its key assumption, which allows the change from the real world probability measure $P$ to the equivalent risk neutral probability measure $Q$, is violated.\\

\ni In the absence of an equivalent risk neutral probability measure, one can still formally form risk neutral prices, which appears to be widely practiced in reality. We will now describe how we formally obtain risk neutral prices under the benchmark approach in a general incomplete market that satisfies Assumptions 1 and 2.\\

\ni First, in the given real-world market model under $P$, set the Radon-Nikodym derivative of the putative risk neutral measure to
$$\label{eq:13}
\Lambda^Q_t = \frac{\ws^1_t}{\ws^1_0}
$$
for $\tin$. This provides a model under the measure $Q$. Second, interpret $Q$ as a probability measure, even though it is most likely not a probability measure under a given realistic model. This step is widely practiced when writing down the risk neutral model dynamics and assuming that its risk neutral measure is an equivalent probability measure, without having formulated the respective model under the real world probability measure. This means that the Radon-Nikodym derivative of the model is formally assumed to be a true martingale, even though this is most likely not the case, since realistic market models are likely to fail the martingale property for the benchmarked savings account, as shown in e.g.~\cite{baldeaux2015pricing}. Third, without checking whether $\Lambda^Q$ forms a martingale, we formally take expectations under the risk neutral model that $Q$ describes, and denote these by $E^Q(\,\cdot\,)$. This leads us to the risk neutral price for a given nonnegative, benchmarked contingent claim $\widehat{H}_\tau$, captured via the risk neutral pricing formula
\beq\label{eq:14}
R^{\widehat{H}_\tau}(t) = S^1_tE^Q\left(\left.\frac{\widehat{H}_\tau}{\ws^1_\tau}\right|\mathcal{F}_t\right)= S^1_tE^Q\left(\left.\frac{H_\tau}{S^1_\tau}\right|\mathcal{F}_t\right)
\eeq
for $\tintau$ with $H_\tau = \widehat{H}_\tau N_\tau$, assuming that the right hand side of (\ref{eq:14}) is finite.\\

\ni By applying Bayes rule, it is straightforward to show that, if $\Lambda^Q_t$ were a martingale, then the benchmarked risk neutral price $\widehat{R}^{\widehat{H}_\tau}(t) = R^{\widehat{H}_\tau}(t)/N_t$ would form a true martingale and would coincide with the benchmarked real world price $\widehat{V}^{\widehat{H}_\tau}(t)$ given in (\ref{eq:2}). However, in general, $\widehat{R}^{\widehat{H}_\tau}(t)$ forms only a local martingale; see Assumption 2, and, thus, by Fatou's Lemma, it forms a supermartingale satisfying (\ref{eq:1'}).\\


\ni As a convenient illustration of risk neutral pricing, consider a zero-coupon bond that pays one unit of the savings account at maturity $T\in(0,\infty)$, i.e. $H_T=S^1_T$ and $\widehat{H}_T=S^1_T/N_T=\ws^1_T$. Then, by (\ref{eq:14}), the risk neutral price for this bond equals $R^{\ws^1_T}(t)=S^1_t$, i.e. one unit of the savings account. Obviously, the nonnegative, benchmarked value of this risk neutral price is
\beq\label{eq:15}
\widehat{R}^{\ws^1_T}(t)=\ws^1_t,
\eeq
which forms a local martingale, and, thus, a supermartingale, where the inequality (\ref{eq:1'}) is satisfied.\\

\ni By Corollary 3, we know that the respective (theoretically possible) minimal zero-coupon bond price $V^{\ws^1_T}(t)$ satisfies (\ref{eq:2}) and the inequality
\beq\label{eq:16}
V^{\ws^1_T}(t)\leq R^{\ws^1_T}(t)=S^1_t
\eeq
for $t\in[0,T]$. Because of (\ref{eq:2}), its benchmarked value $\widehat{V}^{\ws^1_T}(t) = V^{\ws^1_T}(t)/N_t$ forms a true martingale, whereas, $\widehat{R}^{\ws^1_T}(t)$ forms a supermartingale with the same payoff $\widehat{H}_T = \ws^1_T$ at maturity $T$.\\

\ni To link our discussion to the pricing of CAT bonds, let us now consider the payoff of a stylized CAT bond at a fixed maturity date $T$, where the (savings account) discounted payoff $\oline{H}_T=H_T/S^1_T=\widehat{H}_T/\ws^1_T$ is assumed to be independent of the benchmarked savings account value $\ws^1_T$, which can in turn be interpreted as being independent of the financial market. This means that its expected discounted value under $Q$ is assumed equal to that under $P$, i.e. $E^Q(\oline{H}_T|\mathcal{F}_t) = E(\oline{H}_T|\mathcal{F}_t)$. Most importantly, despite the possibility that $\Lambda^Q$ may not be a true martingale, when obtaining a formal risk neutral price, $\Lambda^Q$ is treated as if it were a true martingale, i.e. $E\left(\left.\frac{\Lambda^Q_T}{\Lambda^Q_t}\right|\mathcal{F}_t\right)$ is treated as being equal to 1. Then by equation (\ref{eq:14}), and treating $\Lambda^Q$ as if it were a true martingale, we derive the respective risk neutral price as
\begin{eqnarray*}\label{eq:17}
R^{\widehat{H}_\tau}(t) &=& S^1_tE^Q(\oline{H}_T|\mathcal{F}_t) \nonumber \\
&=& S^1_tE\left(\left.\frac{\Lambda^Q_T}{\Lambda^Q_t}\,\oline{H}_T\right|\mathcal{F}_t\right) \nonumber \\
&=& S^1_tE\left(\left.\frac{\Lambda^Q_T}{\Lambda^Q_t}\right|\mathcal{F}_t\right)E(\oline{H}_T|\mathcal{F}_t) \nonumber \\
&=& S^1_tE(\oline{H}_T|\mathcal{F}_t).
\end{eqnarray*}
On the other hand, the minimal price emerges by formula (\ref{eq:2}) as
\begin{eqnarray*}\label{eq:18}
V^{\widehat{H}_\tau}(t) &=& N_tE(\oline{H}_T\ws^1_T|\mathcal{F}_t) \nonumber \\
&=& N_tE(\ws^1_T|\mathcal{F}_t)E(\oline{H}_T|\mathcal{F}_t) \nonumber \\
    &=& V^{\ws^1_T}(t)E(\oline{H}_T|\mathcal{F}_t).
\end{eqnarray*}
This leads us to the following result:\\

\ni {\bf Corollary 6:} {\it For (savings account) discounted contingent claims $\oline{H}_T = H_T/S^1_T = \widehat{H}_T/\ws^1_T$, delivered at a fixed maturity $T$, that are independent of the benchmarked savings account value $\ws^1_T$, the ratio of the risk neutral price and the minimal price equals the inverse of the (savings account) discounted, minimal price of a zero-coupon bond that pays one unit of the savings account at maturity $T$, i.e.,
\beq\label{eq:19}
\frac{R^{\widehat{H}_T}(t)}{V^{\widehat{H}_T}(t)} = \left(\frac{V^{\ws^1_T}(t)}{S^1_t}\right)^{-1} = \frac{1}{\ovline{V}^{\ws^1_T}(t)}
\eeq
for $t\in[0,T]$.}\\

\ni Then, whatever independent, discounted payoff a risk neutral CAT bond has, the ratio of its risk neutral price to its minimal price is always determined by (\ref{eq:19}), and does not depend on the specification of the discounted payoff. In this case, we have a specific proportionality between risk neutral and minimal prices.

\section{Real World and Risk Neutral Prices under the Minimal Market Model}
\noindent To demonstrate the potential ratio of the risk neutral price to the minimal price in equation (\ref{eq:19}), and to illustrate how loading sensitive pricing works, we introduce the minimal market model (MMM) of~\cite{platen2001minimal}; see also Chapter 13 in~\cite{PlatenHeath06}. The MMM enables us to model a realistic, long-term dynamics of the NP, where the benchmarked savings account turns out to be a strict supermartingale and not a true martingale. \\

\ni Under the MMM, the discounted NP, denoted by $\oline{N}_t = N_t/S^1_t = 1/\ws^1_t$, satisfies the stochastic differential equation (SDE)
\beq\label{eq:20}
d\,\oline{N}_t = \alpha_t\,dt + \sqrt{\alpha_t\,\oline{N}_t}\,dW_t
\eeq
for $\tin$, with initial value $\,\oline{N}_0>0$ and where $W = \{W_t,t\in[0,\infty)\}$ is a standard Brownian motion. Here, $\alpha_t$ denotes the exponential function of time
\beq\label{eq:21}
\alpha_t = \alpha_0\exp\{\eta t\}
\eeq
for $\tin$, with initial value $\alpha_0>0$ and net growth rate $\eta>0$.\\

\ni From the SDE (\ref{eq:20}) and the It\^o formula we can deduce that the benchmarked savings account $\widehat{S}^1_t = 1/\oline{N}_t$ satisfies the SDE
$$\label{eq:22}
d\widehat{S}^1_t = -\sqrt{\alpha_t}\left(\widehat{S}^1_t\right)^{\frac{3}{2}}\,dW_t
$$
for $t\geq 0$. This SDE is driftless, confirming that the benchmarked savings account is a local martingale, as required by Assumption 1. Since one can show that $\oline{N}_t$ is a time-transformed, squared Bessel process of dimension four; see~\cite{RevunYor99}, it follows that it is a strict local martingale and we obtain the expectation
\beq\label{eq:23}
E\left(\left.\frac{\ws^1_T}{\ws^1_t}\right|\mathcal{F}_t\right) = E\left(\left.\frac{\oline{N}_t}{\oline{N}_T}\right|\mathcal{F}_t\right) = 1 - \exp\left\{-\frac{1}{2}\frac{\oline{N}_t}{(\rho_T-\rho_t)}\right\},
\eeq
where
\beq\label{eq:24}
\rho_t = \frac{\alpha_0}{4\eta}(\exp\{\eta t\}-1)
\eeq
for $\tin$; see Section 8.7 in~\cite{PlatenHeath06} for deriving (\ref{eq:23}). Since the right hand side of (\ref{eq:23}) is less than one for $t<T$, the benchmarked savings account is clearly not a true martingale, and only a  strict local martingale. By Fatou's lemma, this makes the nonnegative, benchmarked savings account process $\widehat{S}^1$ a strict supermartingale. As a consequence, the Radon-Nikodym derivative $\Lambda^Q_t = \widehat{S}^1_t/\widehat{S}^1_0$ of the putative risk neutral measure $Q$ is then also a strict supermartingale, and classical risk neutral pricing theory is, therefore, not applicable. However, we can still formally calculate the risk neutral price for a given benchmarked contingent claim (\ref{eq:2}), as described in Section 5. Additionally, we have the minimal price, as well as loading prices for given loading degrees. All of these prices are possible at the same time because their benchmarked prices are local martingales, and the best performing portfolio in the long run is the NP. This NP does remain finite in finite time and, therefore, the model does not permit any economically meaningful arbitrage.\\

\ni It has been shown that the MMM is a reasonably realistic model for the NP of the stock market; see e.g.~\cite{fergusson2014}. The volatility $\sigma_t = \sqrt{\alpha_t/\,\oline{N}_t}$ of the NP in (\ref{eq:20}) models the well-observed leverage effect of equity indices in a natural way. There are key properties of the MMM that support realistic long-term modeling; as discussed in e.g.~\cite{baldeaux2015pricing}. Furthermore, it has been shown in~\cite{PlatenHeath06} and~\cite{PlatenRendek10} that the NP of the stock market can be asymptotically approximated by well-diversified equity indices. This means that a well-diversified, market capitalization weighted stock index can be interpreted as a proxy for the NP, with its dynamics modeled by the MMM.\\


\ni As a convenient illustration of loading pricing, consider, as in (\ref{eq:15}) and (\ref{eq:16}), a zero-coupon bond that pays one unit of the savings account at maturity $T\in(0,\infty)$, i.e. $\oline{H}_T = \widehat{H}_T/\ws^1_T = 1$. The real world pricing formula (\ref{eq:2}) allows us to calculate the (savings account) discounted, minimal zero-coupon bond price $\ovline{V}^{\ws^1_T}(t)$ at time $t<T$, i.e. for $\widehat{H}_T = S^1_T/N_T = \ws^1_T$, via
\beq\label{eq:25}
\ovline{V}^{\ws^1_T}(t) = E\left(\left.\frac{\ws^1_T}{\ws^1_t}\right|\mathcal{F}_t\right).
\eeq By (\ref{eq:23}) and (\ref{eq:25}) we then obtain the expression
\beq\label{eq:26}
\ovline{V}^{\ws^1_T}(t) = 1 - \exp\left\{-\frac{1}{2}\,\frac{\oline{N}_t}{(\rho_T-\rho_t)}\right\}.
\eeq
By (\ref{eq:15}) the respective (savings account) discounted, risk neutral price follows in the form
$$\label{eq:27}
\oline{R}^{\ws^1_T}(t) = 1.
$$
Then $R^{\ws^1_T}(t) = S^1_t$ denotes the risk neutral zero-coupon bond price, and
\beq\label{eq:28}
V^{\ws^1_T}(t) = \ovline{V}^{\ws^1_T}(t)R^{\ws^1_T}(t) = \ovline{V}^{\ws^1_T}(t)S^1_t = S^1_t\left(1 - \exp\left\{-\frac{1}{2}\,\frac{\oline{N}_t}{(\rho_T-\rho_t)}\right\}\right)
\eeq
the minimal zero-coupon bond price.\\

\ni Note that, due to the savings account payout at maturity in $R^{\ws^1_T}$ and $V^{\ws^1_T}$, i.e. $\oline{H}_T = 1$, there is no need to model the interest rate, which in this illustrative example simplifies our modeling and the calculations. There is no impediment to dealing with more complicated contingent claims similarly. Under the MMM, equation (\ref{eq:26}) determines how much smaller the discounted, minimal zero-coupon bond price $\ovline{V}^{\ws^1_T}(t)$ is than the discounted, risk neutral price $\oline{R}^{\ws^1_T}(t) = 1$ for $0\leq t<T<\infty$.\\

\ni The ratio $R^{\widehat{H}_\tau}(t)/V^{\widehat{H}_\tau}(t)$ in (\ref{eq:19}), determines not only how much more expensive the risk neutral zero-coupon bond price turns out to be relative to the minimal zero-coupon bond price, but by Corollary 6, this ratio is the same for all contingent claims with deterministic maturity $T$ and $\oline{H}_T$ independent of $\ws^1_T$. This then also covers the case of a stylized CAT bond, as we demonstrate later on.\\

\ni Let us now visualize for the MMM some of our findings when choosing the S\&P500 as proxy for the NP. We are using monthly data, reconstructed from Global Financial Data, for the period from 1926 to 2015, and present in Figure 1 the logarithm of the savings account, i.e. the discounted S\&P500, which we start in January 1926 with the value 10.0.

\begin{figure}
    \label{Figure 1}
    \begin{center}
        \includegraphics[width=0.75\columnwidth]{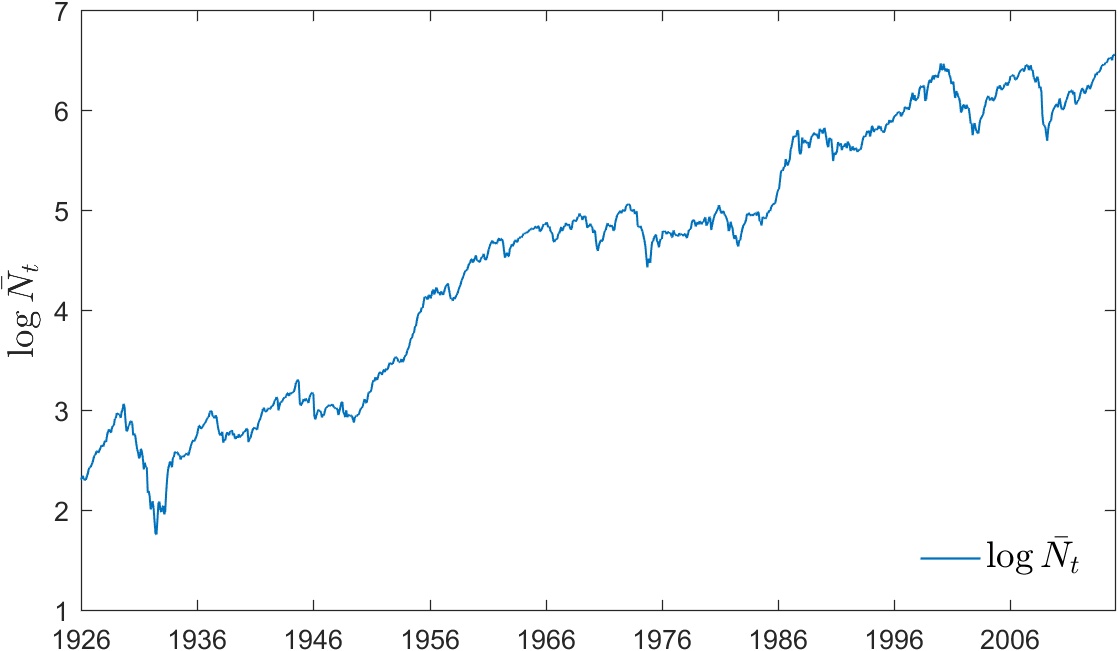}
    \end{center}
    \caption{Logarithm of discounted S\&P500.}
\end{figure}

\ni We fit the parameters of the MMM in the way described in Chapter 13 of~\cite{PlatenHeath06}. For this we note that, by the It\^o formula and  (\ref{eq:20}), the square root of the discounted NP satisfies the SDE
$$
d\sqrt{\oline{N}_t} = \frac{3}{8}\frac{\alpha_t}{\sqrt{\oline{N}_t}}dt + \frac{1}{2}\sqrt{\alpha_t}dW_t.
$$
Thus, by (\ref{eq:21}) and (\ref{eq:24}), the quadratic variation of $\sqrt{\oline{N}_t}$ is given by
\beq\label{eq:30}
\left[ \sqrt{\oline{N}_\cdot} \right]_t = \frac{1}{4}\int^t_0 \alpha_s\,ds = \frac{\alpha_0}{4\eta}(\exp\{\eta t\}-1) = \rho_t
\eeq
for $t\in[0,\infty)$. This means that we observe an estimate for the quantity $\rho_t$ via the quadratic variation of $\left[\sqrt{\oline{N}_\cdot}\right]_t$, with the net growth rate $\eta$ as the key parameter and $\alpha_0$ as some initial parameter.\\

 \ni In Figure 2, we show the observed quadratic variation $\left[ \sqrt{\oline{N}_\cdot} \right]_t$ and, by (\ref{eq:30}), the theoretically calculated $\rho_t$. The quantities $\eta$ and $\alpha_0$ have been fitted by a least squares regression for $\rho_t$, yielding the parameter estimates $\eta = 0.052$ and $\alpha_0 = 0.18$.

 \begin{figure}
     \label{Figure 2}
     \begin{center}
         \includegraphics[width=0.75\columnwidth]{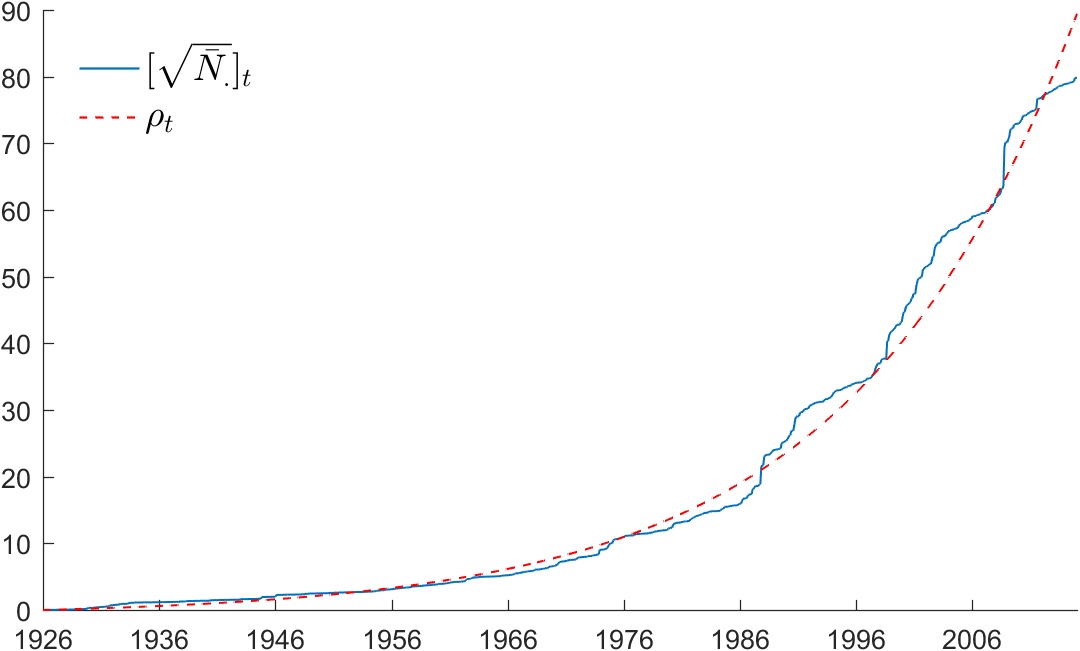}
     \end{center}
     \caption{Quadratic variation of $\sqrt{\oline{N}_t}$ together with $\rho_t$.}
 \end{figure}

\ni The  key structural parameter is the net growth rate $\eta$, which has a clear economic meaning and can be interpreted as the long-term average growth rate of the discounted NP. In our case, where we use the S\&P500 as a proxy for the NP, $\eta$ represents an estimate for the long-term, average net growth rate of the US economy, which means its growth rate above the US interest rate.\\

\ni We now have  the parameters needed to apply the MMM to the available historical data set of the S\&P500. Most interesting is the ratio of the risk neutral price of a zero-coupon bond to the respective minimal price, as described in (\ref{eq:19}). In Figure 3, we show this ratio as a function of time to maturity $(T-t)$, where we fix the maturity as the final date of our time series in January, 2015. By Corollary 6, this is also the ratio of the risk neutral price to that of the respective minimal price of a contingent claim when its discounted value is independent of the benchmarked savings account, which includes the case of stylized CAT bonds.\\

\begin{figure}
    \label{Figure 3}
    \begin{center}
        \includegraphics[width=0.75\columnwidth]{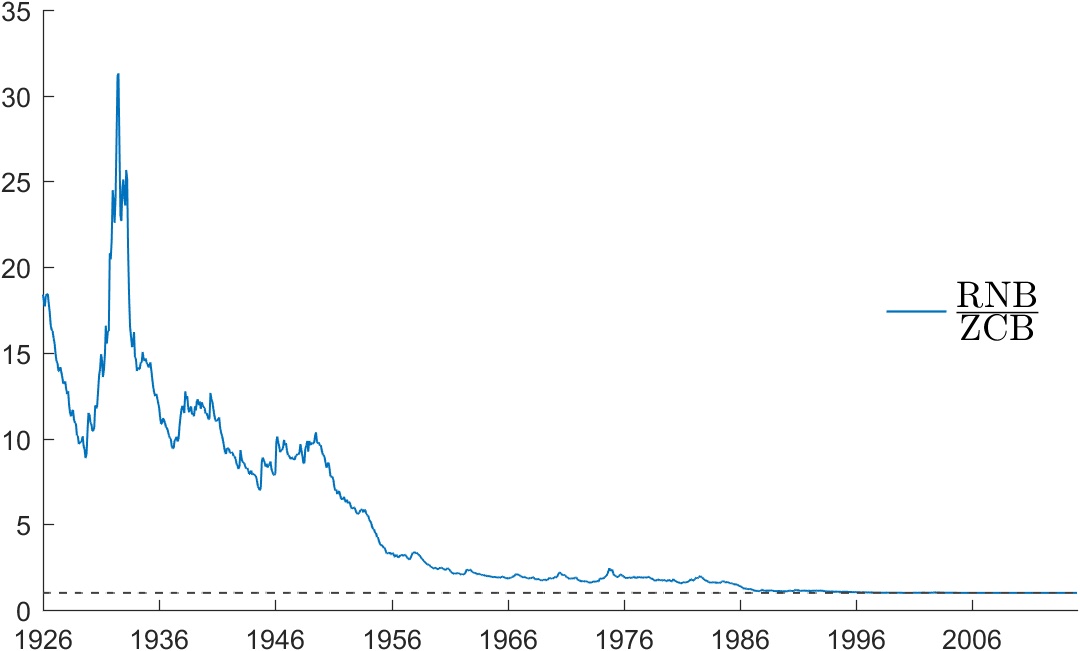}
    \end{center}
    \caption{Ratio of the (formally obtained) risk neutral price to the (theoretically possible) minimal price of a zero-coupon bond as a function of time with fixed maturity.}
\end{figure}

\ni One notes that the risk neutral price is about double that of the minimal price for a time to maturity of roughly 30 years. It is important to observe that for maturities of up to 10 years there is not much difference between the respective minimal and risk neutral prices. This reflects the fact that the benchmarked savings account, as a local martingale, behaves like a martingale in the short term. If one were to employ a better performing proxy for the NP than the S\&P500, e.g. the one described in~\cite{PlatenRendek10}, then one could expect a higher net growth rate $\eta$ and, therefore, larger differences between minimal and risk neutral prices.\\

\ni To illustrate loading pricing further, let us now examine a stylized CAT bond. Let, $\oline{H}_T = \mathbbm{1}_{\{\xi\in[0,T]\}}\geq 0$ be the agreed discounted payout of one unit of the savings account at maturity $T$, in the event that a specified catastrophe has occurred at some stopping time $\xi>0$ prior to $T$. Initiation of the contract is at $t=0$, and we assume that the  time $\xi$ of the catastrophe is independent of the benchmarked savings account value $\ws^1_T$. This yields the benchmarked payoff $\widehat{H}_T = \oline{H}_T\ws^1_T = \mathbbm{1}_{\{\xi\in[0,T]\}}\ws^1_T$ and the respective conditional expectation
\begin{eqnarray}\label{eq:31}
E(\widehat{H}_T|\mf_t) &=& E(\mathbbm{1}_{\{\xi\in[0,T]\}}|\mf_t)E(\ws^1_T|\mf_t) \nonumber \\
&=& \oline{H}_t\widehat{V}^{\ws^1_T}(t),
\end{eqnarray}
where the real world conditional expectation (conditional probability)
$$
\oline{H}_t = E(\mathbbm{1}_{\{\xi\in[0,T]\}}|\mf_t) = P(\xi\in[0,T]|\mf_t)
$$
forms a martingale.\\

\ni In the case of a stylized CAT bond that pays one unit $S^1_T$ of the savings account at time $T$ in the event that the insured catastrophe has occurred at some time $\xi$ prior to $T$ or nothing otherwise, the respective minimal price is, by (\ref{eq:2}), (\ref{eq:26}) and (\ref{eq:31}) of the form
\begin{eqnarray}\label{eq:32}
V^{\widehat{H}_T}(t) &=& N_t E(\widehat{H}_T|\mf_t) \nonumber \\
&=& \oline{H}_tV^{\ws^1_T}(t) \nonumber \\
&=& \oline{H}_t S^1_t\left(1-\exp\left\{-\frac{1}{2}\,\frac{\oline{N}_t}{(\rho_T - \rho_t)}\right\}\right).
\end{eqnarray}
This means that the minimal price of the stylized CAT bond equals the product of the conditional probability that the catastrophe occurs prior to $T$ and the minimal bond price.\\

\ni By Corollary 6, the corresponding risk neutral price has the form
\beq\label{eq:33}
R^{\widehat{H}_T}(t) =\oline{H}_tS^1_t
\eeq
for $t\in[0,T]$. This means that $R^{\widehat{H}_T}(t)$ equals the product of the previously mentioned conditional probability and the savings account value, the latter representing the risk neutral bond price.\\

\ni For a given loading degree $L^{\widehat{H}_T} = \{L^{\widehat{H}_T}_t,t\in[0,T]\}$, we get the respective loading price from (\ref{eq:7}), (\ref{eq:32}) and (\ref{eq:33}) as
$$\label{eq:34}
B^{\widehat{H}_T}(t) = \oline{H}_tS^1_t\left(1-(1-L^{\widehat{H}_T}_t)\exp\left\{-\frac{1}{2}\,\frac{\oline{N}_t}{(\rho_T - \rho_t)}\right\}\right)
$$
at time $t\in[0,T)$. One notes that the payout $S^1_T\mathbbm{1}_{\{\xi\in[0,T]\}}$ is delivered at maturity $T$ by the loading price process $B^{\widehat{H}_T}$, and its price is always larger than or equal to the minimal price.\\

\ni A simple case for the loading price of a stylized CAT bond is the one for a constant loading degree $L^{\widehat{H}_T}_t = L^{\widehat{H}_T}_0$. More generally, the loading degree can be, e.g., a local martingale, independent of the event time $\xi$ and the benchmarked savings account. This local martingale could then model the ongoing search by the market for an appropriate CAT bond price. This search would most likely be driven by competition between issuers and by supply-and-demand. It is extremely important to note that the prices of contracts or derivatives may involve a new, independent source of uncertainty, i.e. the potential randomness of the loading degree. Obviously, this potential uncertainty can even apply to fully-replicable contingent claims. This shows that there is substantial flexibility in possible price formation under the benchmark approach, without generating any economically meaningful arbitrage. Through competition and supply-and-demand one can probably expect that market prices of similar contracts are characterized by similar loading degrees. By extracting the respective typical loading degree from market prices for a range of insurance contracts, one may be able to value new contracts that are similar to the existing ones in a market-consistent fashion by using this estimated typical loading degree. Since the current paper is more conceptual in nature, we will extract loading degrees from insurance data in forthcoming work.

\section{Hedging under the MMM}
\ni In this section we discuss hedging strategies for zero-coupon bonds. The asset allocation strategy that generates the (formally obtained) risk neutral zero-coupon bond invests its wealth fully in the savings account $S^1$. This is a simple buy-and-hold strategy.\\

\ni Let us describe the dynamic asset allocation, i.e. the hedging strategy, which generates the (theoretically possible) minimal zero-coupon bond. Denote as before by $\ovline{V}^{\ws^1_T}(t)$, the discounted minimal zero-coupon bond price, and by $\oline{R}^{\ws^1_T}(t)=1$, the discounted risk neutral zero-coupon bond price. Then it follows from (\ref{eq:26}) that the hedge ratio for  holding units in the discounted NP $\oline{N}_t = N_t/S^1_t$ at time $t\in[0,T)$ for hedging the minimal zero-coupon bond is
\begin{eqnarray*}\label{eq:35}
\delta_t &=& \frac{\partial\, \ovline{V}^{\ws^1_T}(t)}{\partial\, \oline{N}_t} \nonumber \\
&=&  \frac{2\eta}{\alpha(\exp\{\eta T\}-\exp\{\eta t\})}\exp\left\{\frac{-2\eta\,\oline{N}_t}{\alpha(\exp\{\eta T\}-\exp\{\eta t\})}\right\}.
\end{eqnarray*}
The self-financing hedge portfolio  for $\ovline{V}^{\ws^1_T}(t)$ holds the remainder of the value of the minimal zero-coupon bond in the savings account. For illustration, we consider hedging the minimal zero-coupon bond with maturity in January, 2015, using the S\&P500 as proxy for the NP. The fraction of wealth held in the S\&P500 as a function of time is then given as
\beq\label{eq:36}
\pi_t = \frac{\delta_t\,\oline{N}_t}{\ovline{V}^{\ws^1_T}(t)}
\eeq
and shown in Figure 4.\\

\begin{figure}
    \label{Figure 4}
    \begin{center}
        \includegraphics[width=0.75\columnwidth]{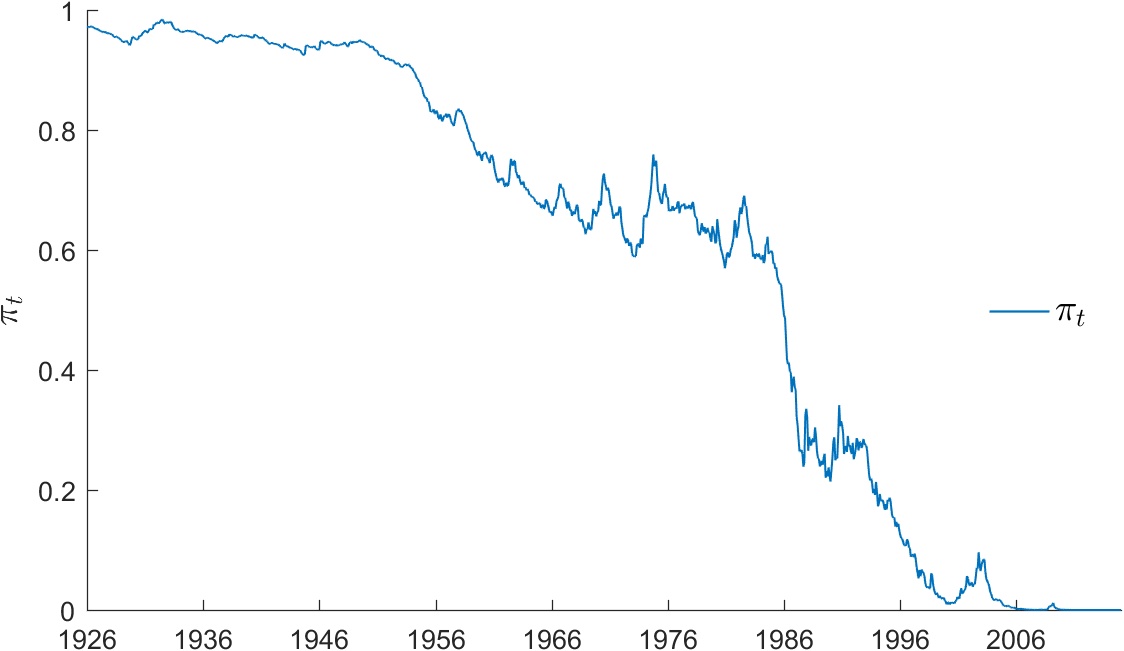}
    \end{center}
    \caption{Fraction invested in the NP (S\&P500) for the (theoretically possible) minimal zero-coupon bond with maturity in January, 2015.}
\end{figure}

\ni We note that for the initial 30 years this hedge portfolio invests primarily in the NP. According to formula (\ref{eq:36}) it slides over later, more and more, into the savings account. When aiming for a savings account unit payout at maturity, one can interpret the above strategy as making common financial planning advice rigorous. Recall that the widely followed financial planning advice initially suggests investing in risky securities and later more and more in fixed income securities when nearing retirement, as is shown in Figure 4 for the least expensive hedge under the given model.\\

\ni For further illustration, we show in Figure 5 the logarithm of the discounted self-financing hedge portfolio with monthly reallocations for the discounted, minimal zero-coupon bond price process. It differs so little from the discounted bond price process that one would not see any difference in the plots. Obviously, the logarithm of the discounted, risk neutral bond has a constant value of zero. One notes that the discounted minimal zero-coupon bond price is initially considerably smaller than the risk neutral bond price. More precisely, as already visible from Figure 3, for the 89-year, zero-coupon bond, the risk neutral bond price in 1926 is about 18 times larger than the minimal zero-coupon bond price.\\

\ni It is straightforward to form a loading price process with a given constant loading degree. In Figure 5, we also show the logarithm of the discounted loading price of a zero-coupon bond with a constant loading degree of $L^{\ws^1_T}_t = 0.3,\;0\leq t\leq T$.\\

\begin{figure}
    \label{Figure 5}
    \begin{center}
        \includegraphics[width=0.75\columnwidth]{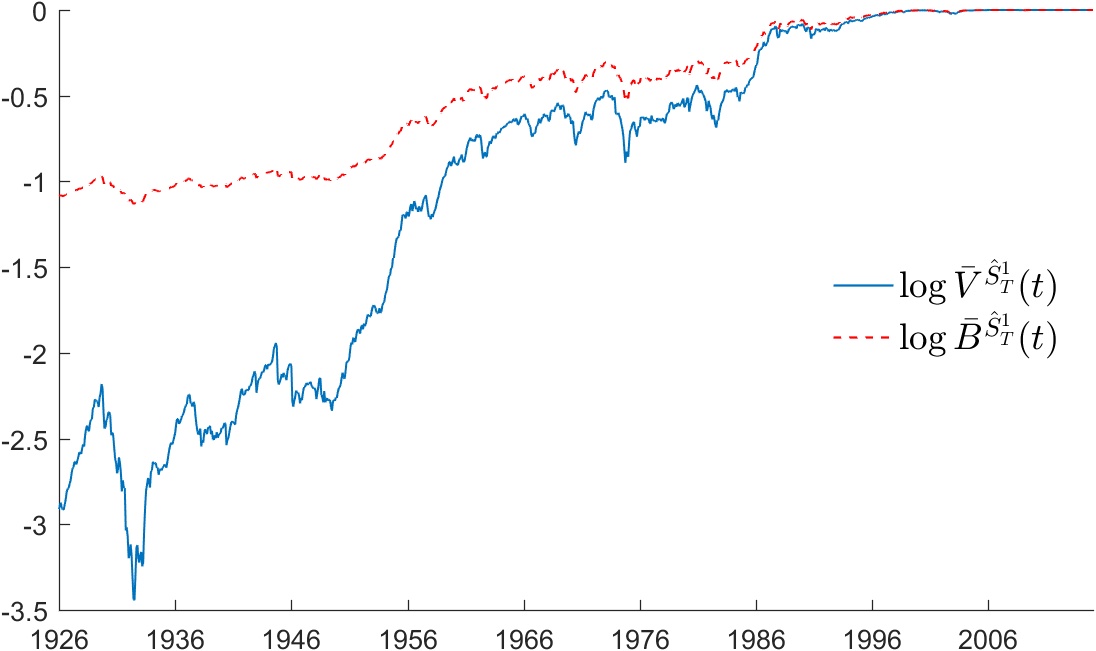}
    \end{center}
    \caption{Logarithms of the monthly-rebalanced hedge portfolio for the discounted, (theoretically possible) minimal zero-coupon bond price, and the discounted loading price process for a constant loading degree $L^{\ws^1_T}_t = 0.3$.}
\end{figure}

\ni The minimal price and with it all loading prices are consistent with the NP in the sense that they form local martingales when benchmarked. This means, if one forms the NP by using all securities, contracts and derivatives traded in the market including loading price processes, then it still emerges as the same NP that we started with. There exists, therefore, no strictly positive portfolio that outperforms pathwise the NP in the long run. This means that there is no strictly positive portfolio that generates infinite wealth out of finite initial capital in finite time. Consequently, we have no economically meaningful arbitrage. Thus, one can exploit the concept of loading pricing to analyze market prices and form market consistent prices for similar new contracts. By estimating or assuming some loading degree for a range of contracts that may be rather illiquid or difficult to evaluate we can, by using respective loading prices, evaluate even a portfolio of insurance contracts, e.g., for regulatory purposes.

\section{Benchmarked Risk Minimization}
\ni In the previous section we considered the hedging of fully-replicable contingent claims in the case of zero-coupon bonds. Without being able to go deeper into the important question of risk minimization, let us now briefly discuss the hedging of not-fully-replicable claims. We use, as an example, the hedging of stylized CAT bonds, characterized previously by a discounted payoff $\,\oline{H}_T$, independent of the benchmarked savings account value $\ws^1_T$. \\

\ni We first aim to price and hedge in the least expensive manner and with minimal possible fluctuations of the, so called, benchmarked profit-and-loss. This we achieve by applying the concept of benchmarked risk minimization; introduced in~\cite{du2016benchmarked}. The hedge portfolio at time $t=0$ is formed with a value equal to the minimal price $V^{\,\oline{H}_T}(0) = \oline{H}_0V^{\ws^1_T}(0)$; see (\ref{eq:32}). Via the It\^o formula and (\ref{eq:32}), we obtain the following martingale representation for the benchmarked contingent claim
\begin{eqnarray}\label{eq:37}
\widehat{H}_T &=& \widehat{V}^{\widehat{H}_T}(T) = \oline{H}_T\widehat{V}^{\ws^1_T}(T) \nonumber \\
&=& \oline{H}_t\widehat{V}^{\ws^1_T}(t) + \int^T_t \oline{H}_{s-}d\widehat{V}^{\ws^1_T}(s) + \int^T_t \widehat{V}^{\ws^1_T}(s-)d\,\oline{H}_s,
\end{eqnarray}
with $\,\oline{H}_t = E(\oline{H}_T|\mf_t)$ and
$$
\widehat{V}^{\widehat{H}_T}(t) = \oline{H}_t\widehat{V}^{\ws^1_T}(t) = E(\widehat{H}_T|\mf_t)
$$
for $t\in[0,T]$. Here we have, by (\ref{eq:36}) and (\ref{eq:28}),
\beq\label{eq:38}
d\widehat{V}^{\ws^1_T}(t) = \widehat{V}^{\ws^1_T}(t-)(1-\pi_{t-})\frac{d\ws^1_t}{\ws^1_{t-}} = \ovline{V}^{\ws^1_T}(t-)(1-\pi_{t-})d\ws^1_t,
\eeq
with $\pi_t$ given in (\ref{eq:36}) denoting the fraction of wealth invested in the NP for hedging the minimal bond $V^{\ws^1_T}$ at time $t\in[0,T]$. By holding
$$\label{eq:39}
\delta_t^1 = \oline{H}_{t-}\ovline{V}^{\ws^1_T}(t-)(1-\pi_{t-})
$$
units of the savings account in the hedge portfolio at time $t$, according to~\cite{du2016benchmarked}, and because of (\ref{eq:37}) and (\ref{eq:38}), the benchmarked profit-and-loss at time $t$ equals
\beq\label{eq:40}
\widehat{C}^{\widehat{H}_T}_t = \widehat{V}^{\widehat{H}_T}(t) - \int^t_0 \delta^1_{s-}d\ws^1_s - \widehat{V}^{\widehat{H}_T}(0) = \int^t_0 \widehat{V}^{\ws^1_T}(s-)d\oline{H}_s.
\eeq
It is required under benchmarked risk minimization; see~\cite{du2016benchmarked}, that $\widehat{C}^{\widehat{H}_T}$ is a local martingale, orthogonal to all benchmarked traded wealth. In particular, it has to be orthogonal to $\ws^1$. Its products with benchmarked primary security accounts need to satisfy SDEs with zero drifts. In this sense, through benchmarked risk minimization, the fluctuations of the benchmarked profit-and-loss process are minimized; see~\cite{du2016benchmarked} for further information. The amount $V^{\widehat{H}_T}(0) = N_0E(\widehat{H}_T|\mf_0)$ is the initial amount needed for this hedge with $\delta^1_0$ units held in the savings account at time $t=0$. The number of units
\beq\label{eq:41}
\delta^*_t = \widehat{V}^{\widehat{H}_T}(t) - \delta_t^1\ws^1_t
\eeq
are then held in the NP at time $t\in[0,T]$. We note from (\ref{eq:38}), (\ref{eq:40}) and (\ref{eq:41}) that $\delta^*_t = \widehat{C}^{\widehat{H}_T}_t$ equals the benchmarked profit-and-loss $\widehat{C}^{\widehat{H}_T}_t$. This emphasizes the fact that we do not have here a self-financing hedge portfolio for a CAT bond and one can interpret $\delta^*_t$ as the additional holdings in the NP needed to deliver the benchmarked contingent claim $\widehat{H}_T$ at maturity $T$.\\

\ni An additional interesting fact comes to light when we compare the classical risk neutral hedging strategy for a discounted, non-hedgeable claim with the benchmarked risk minimizing approach. The classical risk neutral hedging; see e.g.~\cite{Schweizer99} or~\cite{moller2001risk}, charges the risk neutral price at initiation of the contract and invests this amount in the savings account until maturity. Evolving information is ignored here until maturity. By (\ref{eq:40}) we see that evolving information about the non-hedgeable part of the claim, encapsulated here by the martingale $\oline{H}_t$, is taken into account. In practice, insurers take evolving information about catastrophes into account, e.g. by raising extra capital after severe catastrophes. Classical risk minimization ignores evolving information, whereas benchmarked risk minimization incorporates it in a natural manner.\\

\ni With the above hedging strategy we have characterized the least expensive hedge with minimal fluctuations of the benchmarked profit-and-loss that delivers in the mean the targeted benchmarked contingent claim. For the large book of contracts of an insurance company with an increasing number of (sufficiently) independent, benchmarked profits-and-losses, generated by respective contracts, one can see that, due to the Law of Large Numbers, the benchmarked total profit-and-loss converges almost surely to zero; see~\cite{du2016benchmarked}. In this sense, there is asymptotically no residual, unhedgeable risk remaining. The contingent claims are then hedged in the least expensive manner and the unhedgeable uncertainty is asymptotically removed by diversification. We will discuss in the next section how this can be generalized to involve loading pricing.

\section{Risk Minimization under Loading}
\ni The proposed loading pricing allows the issuer of a CAT bond to ask systematically for a higher price than the minimal one. In a competitive market the loading degree emerges from market prices that balance the need for insurance companies to attract customers in competition with other insurance companies, and the need to build up and maintain capital reserves to avoid ruin, caused by exposure to unhedged risk. \\

\ni We emphasized already that the loading degree process has to be chosen appropriately to avoid economically meaningful arbitrage and pointed out that there is theoretically substantial scope for possible loading degree processes.\\

\ni On the other hand, one would expect for economic reasons that prices in a competitive market for similar products would tend to get similar loading degrees, which hints at the potential existence of a possibly similar loading degree for a range of similar products. An issuer of CAT bonds can invest in a capital reserve the extra amount received via the loading price above the minimal price. The aggregate of this capital reserve sits on top of the sum of the minimal prices needed to deliver on average the targeted benchmarked payoffs in the insurer's book. The above benchmarked risk minimizing hedging strategy invests all extra funds in the NP. \\

\ni When performing risk minimization under loading, we propose, therefore, that capital reserves should also be invested in the NP here, which benefits the growth of the company's reserves in the long run. The benchmarked profit-and-loss should also be a local martingale here, orthogonal to benchmarked traded wealth, as was the case under benchmarked risk minimization. The loading price process should remain the same as discussed in previous sections. The capital reserve collected and accumulated from charging loading prices then represents the buffer needed to prevent ruin of the insurance company. By performing this type of loading risk minimization, one minimizes the fluctuations of the benchmarked profit-and-loss, hedges the hedgeable part of contingent claims under the minimal possible expense, and consistently charges respective loading prices. Diversification of the benchmarked profits-and-losses of the contracts issued by an insurer asymptotically removes the nonhedgeable uncertainty, which has always been a prudent strategy. What the proposed risk minimization under loading does is to make the intuitive and experientially developed investment and risk management processes in insurance companies more rigorous.\\

\ni Current practice and regulators may suggest slightly different strategies, in particular, strategies that may cause more and more problems by requiring capital reserves to be held in fixed income securities, the savings account, which in many countries currently have very low or even negative returns. Our loading pricing concept together with its hedging strategy allows the creation of insurance products that provide the insured with higher return on investment and the insurer with a less expensive hedging strategy than permitted under classical assumptions. Very importantly, the capital reserves are invested in the NP, the best performing portfolio in the long-run, and not in the savings account.\\

\ni Similarly to benchmarked risk minimization, let us now analyze the benchmarked profit-and-loss for the stylized CAT bonds of Sections 5 to 8, when using the loading price process we established in Section 4. The key question is how to hedge such a stylized CAT bond when priced via a loading degree process such that the benchmarked profit-and-loss exhibits minimal fluctuations.\\

\ni For a given benchmarked contingent claim $\widehat{H}_T$, delivered at a deterministic maturity date $T$ and priced using the loading degree process $L^{\widehat{H}_T}$ (which is a local martingale orthogonal to $\widehat{R}^{\widehat{H}_T} - \widehat{V}^{\widehat{H}_T}$), we assume the benchmarked loading price process $\widehat{B}^{\widehat{H}_T}$ given by (\ref{eq:9'}) and denote by $\delta^{L,1}_t$ the number of units of the savings account held at time $t\in[0,T]$ in the respective strategy that makes the benchmarked profit-and-loss a local martingale, orthogonal to benchmarked traded wealth.\\

\ni The respective benchmarked profit-and-loss under loading risk minimization is defined by the equation
$$
\widehat{C}^{\widehat{H}_T}_t = \widehat{B}^{\widehat{H}_T}(t) - \int^t_0 \delta^{L,1}_{s-}\,d\ws^1_s - \widehat{V}^{\widehat{H}_T}(0),
$$
for $t\in[0,T]$, where we now allow some positive initial benchmarked profit-and-loss $\widehat{C}^{\widehat{H}_T}_0 = \widehat{B}^{\widehat{H}_T}(0) - \widehat{V}^{\widehat{H}_T}(0)\geq 0$. Using equation (\ref{eq:9'}), together with the It\^o formula and the previous notation and properties of the quantities involved, we obtain
\begin{eqnarray*}
\widehat{C}^{\widehat{H}_T}_t &=& \widehat{C}^{\widehat{H}_T}_0 + \int^t_0 (1-L^{\widehat{H}_T}_{s-})\,d\widehat{V}^{\widehat{H}_T}(s) + \int^t_0 L^{\widehat{H}_T}_{s-}\,d\widehat{R}^{\widehat{H}_T}(s) +
\int^t_0 \left(\widehat{R}^{\widehat{H}_T}(s-) - \widehat{V}^{\widehat{H}_T}(s-)\right)dL^{\widehat{H}_T}_s \\
&& - \int^t_0 \delta^{L,1}_{s-}\,d\ws^1_s \\
&=& \widehat{C}^{\widehat{H}_T}_0 + \int^t_0 \left((1-L^{\widehat{H}_T}_{s-})\,\oline{H}_{s-}\ovline{V}^{\ws^1_T}(s-)(1-\pi_{s-}) + L^{\widehat{H}_T}_{s-}\,\oline{H}_{s-} - \delta^{L,1}_{s-}\right)d\ws^1_s \\
&& + \int^t_0 \left(L^{\widehat{H}_T}_{s-}\ws^1_{s-} + (1-L^{\widehat{H}_T}_{s-})\widehat{V}^{\ws^1_T}(s-)\right)d\,\oline{H}_s + \int^t_0 \left(\widehat{R}^{\widehat{H}_T}(s-) - \widehat{V}^{\widehat{H}_T}(s-)\right)dL^{\widehat{H}_T}_s.
\end{eqnarray*}
Clearly, $\widehat{C}^{\widehat{H}_T}$ is a local martingale, because all integrators are local martingales. Its fluctuations can be minimized by first setting the number of units held in the savings account to
\beq\label{eq:42}
\delta^{L,1}_t = \oline{H}_t\left((1-L^{\widehat{H}_T}_t)\ovline{V}^{\ws^1_T}(t)(1-\pi_t) + L^{\widehat{H}_T}_t\right)
\eeq
and second by making the loading degree process a constant. It is important to note that avoiding a fluctuating loading degree eliminates removable uncertainty from the loading price process, which we summarise as follows:\\

\noindent {\bf Corollary 7:} {\it A constant loading degree process together with the choice (\ref{eq:42}) yield the ``smallest" fluctuations of benchmarked profit-and-loss under loading risk minimization, which takes the form
$$
\widehat{C}^{\widehat{H}_T}_t = \widehat{C}^{\widehat{H}_T}_0 + \int^t_0\left(L^{\widehat{H}_T}_{s-}\ws^1_{s-} + (1-L^{\widehat{H}_T}_{s-})\widehat{V}^{\ws^1_T}(s-)\right)d\,\oline{H}_{s-}
$$
for $t\in[0,T]$.}\\

\ni This insight opens the search for some empirical loading degrees that the market most likely forms for a range of similar insurance products.\\

\ni In~\cite{gurtler2014impact}, it has been shown empirically that some increase in prices (loadings) of CAT bonds typically arise after financial crises and natural catastrophes. Our loading pricing is able to model such observations consistently, as will be described in forthcoming work. There we will provide empirical evidence for loading degrees in zero-coupon bonds, various CAT bonds, and other insurance-type contracts.

\section*{References}
\bibliographystyle{plainnat}
\bibliography{biblio}

\end{document}